\documentclass[conference]{IEEEtran}
\IEEEoverridecommandlockouts
\usepackage[top=0.75in, bottom=0.75in, left=0.63in, right=0.63in]{geometry}
\usepackage{cite}
\usepackage{amsmath}
\usepackage{times}
\usepackage{epsfig}
\usepackage{graphicx}
\usepackage{subfig}
\usepackage{amsmath}
\usepackage{amssymb}
\usepackage{tabularray}
\usepackage{tikz}
\usepackage{multirow}
\usepackage{booktabs}
\usepackage[ruled,vlined]{algorithm2e}  
\usepackage{stmaryrd}

\usepackage{caption}  
\usepackage{longtable} 
\usepackage{float} 
\usepackage{url}
\usepackage{subcaption}
\usepackage{xurl}
\usepackage{afterpage}
\usepackage{comment}

\begin{document}
%
\title{Bruxism Recognition via Wireless Signal}
%
%
%

\author{Qiankai~Shen,
        Yuanhao~Cui,
        Jie~Yang,
        Xiaojun~Jing,
        Zhiyong~Feng,
        Shi~Jin,
\thanks{
    Qiankai Shen, Yuanhao Cui, Xiaojun Jing, Zhiyong Feng are with the School of Information and Communication Engineering, Beijing University of Posts and Telecommunications, Beijing 100876, China (sqk140058@bupt.edu.cn; yuanhao.cui@bupt.edu.cn;jxiaojun@bupt.edu.cn;fengzy@bupt.edu.cn).
        
    Jie Yang is with the School of Control of Complex Systems of Engineering, Southeast University, Nanjing 210096, China (yangjie@seu.edu.cn).
    
    Shi Jin is with the School of Information Science and Engineering, Southeast University, Nanjing 210096, China (jinshi@seu.edu.cn).
    
    The corresponding author is Yuanhao Cui.
    }
        }

%
%

\markboth{Journal of \LaTeX\ Class Files,~Vol.~14, No.~8, August~2015}%
{Shell \MakeLowercase{\textit{et al.}}: Bare Demo of IEEEtran.cls for IEEE Journals}
%



\maketitle

\begin{abstract}
Bruxism is an oromandibular movement disorder involving teeth grinding and clenching, which severely impairs sleep quality and dental health. However, its diagnosis remains challenging, as existing methods often cause discomfort or compromise user privacy. To address these limitations, we establish a contactless bruxism recognition system based on millimeter-wave radar. First, we analyzed the potential impact of the movement patterns of teeth grinding on radar echo signals. Based on this analysis, 11 features were extracted. Subsequently, using these features, we performed classification with a Random Forest model on the dataset constructed via millimeter-wave radar.
Experimental results demonstrate that the proposed method achieves an accuracy of 96.1\% on the test set, with precision, recall, and F1-score all remaining at a relatively high level. This study validates the effectiveness of millimeter-wave radar for SB recognition, providing a non-invasive and privacy-friendly alternative to existing recognition techniques. Future research will focus on enhancing the robustness of the method across diverse populations and environments, as well as striving to mitigate the interference of other facial micro-movements on teeth grinding recognition.
\end{abstract}


%
\IEEEpeerreviewmaketitle

\section{Introduction}
Bruxism is an oromandibular movement disorder closely related to sleep physiology, typically occurring during the night and potentially disrupting the normal rhythm and architecture of sleep. Affecting approximately 31.4\% of the population, it represents a clinically significant condition rather than a rare   \cite{population}. Unlike its awake counterpart, bruxism during sleep typically occurs without patient awareness, complicating both diagnosis and management. Repeated nocturnal jaw activity can lead to progressive dental wear, restoration damage, tooth hypersensitivity, temporomandibular joint discomfort, and even chronic pain and sleep disruption that ultimately impair overall oral and sleep health \cite{impact}.

Over the past decades, researchers have explored diverse technologies to detect and characterize bruxism. Among them, surface electromyography (sEMG) has emerged as the predominant technique, owing to its ability to capture muscle activity from the masseter, a key driver of jaw movement. Building on this, Heyat et al. employed EMG-based monitoring in conjunction with machine learning to enable automatic bruxism classification \cite{Heyat}. Extending this work, Gul et al. examined both the masseter and temporalis muscles under varying sleep postures, revealing that signal acquisition sites and body position strongly influence classification accuracy up to 93.3\% \cite{Gul}. These findings emphasize the critical role of methodological refinements in enhancing recognition reliability.

To mitigate the limitations of wired EMG setups, alternative intraoral sensing methods have been explored. Aoki et al. developed a splint-embedded force recognition device with promising recognition performance in home settings, although it exhibited deviations in force estimation relative to EMG \cite{Aoki}. In addition, Koide et al. introduced a miniaturized wearable EMG system capable of continuous monitoring for 28 nights, demonstrating the feasibility of long-term, real-world deployment \cite{Koide}.

In parallel with contact-based techniques, contactless sensing methods have been investigated for bruxism recognition, with audio-based approaches receiving particular attention \cite{microphone}. Peruzzi et al. achieved promising results using microphone-based recognition, yet subsequent studies emphasized that environmental noise and microphone placement critically affect reliability \cite{peruzzi}. To improve robustness, Maoddi et al. compared intraoral force sensing with portable EMG and confirmed strong signal correlations, suggesting that multimodal fusion could improve recognition accuracy \cite{maoddi}. Further advancing this trend, Kostka et al. proposed the integration of accelerometer, EMG, and heart rate features to enable real-time multimodal recognition of bruxism activity \cite{kostka}.

Despite these advances, current recognition methods face persistent limitations. Existing invasive systems cause discomfort during extended wear, reducing long-term compliance, while contactless sensing methods, such as audio-based recognition, raise privacy concerns. These challenges highlight the ongoing requirement for contactless, comfortable, and privacy-preserving solutions, which motivates the recent exploration of wireless and radar-based sensing for bruxism recognition. 

In response to these requirements, we constructa radar-based bruxism recognition system leveraging millimeter-wave (mmWave) sensing., which able to provide a new solution for long-term and affordable bruxism recognition. The system is designed to enable continuous, contactless monitoring without compromising user comfort or privacy, offering a promising alternative for long-term and cost-effective deployment. Furthermore, the proposed method holds potential for early-stage bruxism recognition, which could facilitate timely intervention and improved clinical outcomes. The main contributions of this work are as follows:
\begin{itemize}
    
\item We introduced a radar-based bruxism recognition system that is contactless, comfortable, and privacy-preserving. By leveraging a 60 GHz millimeter-wave radar, we capture subtle facial micro-movements associated with bruxism without requiring physical contact or audio input, addressing the key limitations of existing EMG- and sound-based methods.

\item We developed an end-to-end radar signal processing pipeline tailored for facial jaw-motion extraction. We first apply Range FFT in the fast-time domain to transform the raw radar signals into range profiles. Then, we perform incoherent accumulation across the slow time domain to localize the range bin corresponding to the face area. We finally extract the phase from the selected bin and compute phase differences to capture fine-grained motion features relevant to bruxism activity.

\item We construct a dedicated jaw-motion dataset and validate the system through real-world experiments. The dataset comprises 180 samples, each consisting of a five-second radar echo recorded in an indoor environment. We then train a random forest classifier on the collected data, achieving a test accuracy of 96.1\%. To further evaluate the system's effectiveness, we compare its performance against EMG- and audio-based baselines, demonstrating superior reliability and broader applicability of the proposed radar-based solution.
\end{itemize}

        
    

\section{Bruxism and Its Signal Characteristics}
Bruxism is a movement disorder characterized by grinding and clenching of the teeth, and this study focuses primarily on the identification of grinding teeth. To evaluate the influence of bruxism on radar signals, it is essential first to characterize the movement patterns associated with grinding teeth. This activity mainly involves the masticatory muscles, most notably the masticator and temporalis, whose contractions cause localized bulging of facial tissue. When these muscles contract, both exhibit localized bulging. Simultaneously, the interaction of the upper and lower teeth produces sliding motions that are chiefly responsible for mechanical dental damage. Building on these characteristic motion dynamics, we examine how teeth grinding alters radar echo signals within the facial range domain, where its effects are primarily reflected as phase variations. Because radar phase values naturally drift over time, the analysis focuses on phase differences, which more effectively capture fine motion fluctuations.

\begin{figure}[!t]
\centerline{\includegraphics[width=1\linewidth]{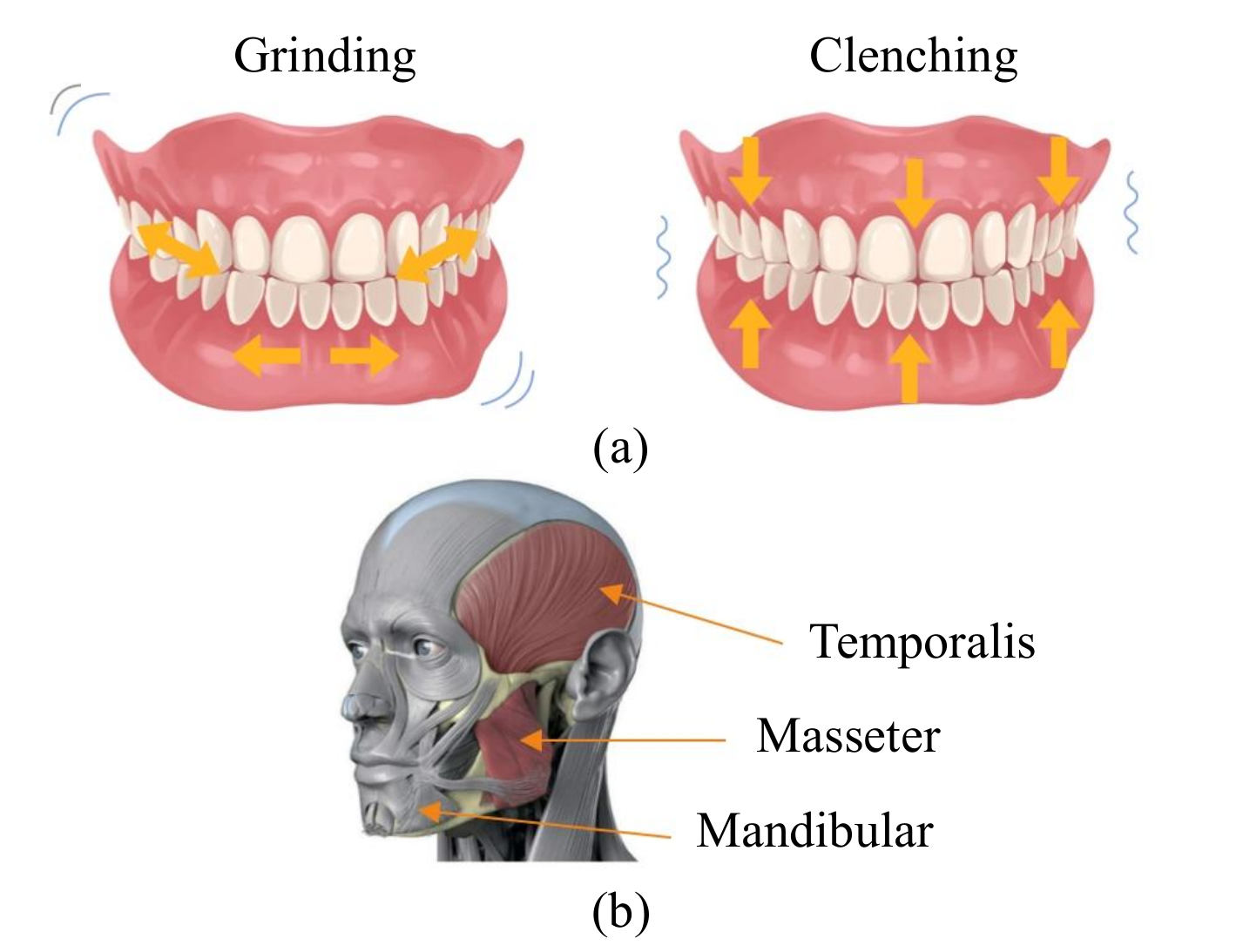}}
\caption{Types of bruxsim and the involved muscles and regions in grinding. (a) Types of bruxsim(grinding and clenching).  (b) the involved muscles and regions in grinding.}
\label{bruxism}
\end{figure}
\subsection{The Signal Effect of Mandibular Oscillations}
During bruxism episodes, the mandible exhibits high-frequency micro-oscillations, typically ranging from 5 to 10 Hz. These vibrations are driven by repetitive stick-slip dynamics on the occlusal surfaces, where static and kinetic friction alternate rapidly. As a result, micro-vibrations on the skin surface caused by mandibular motion become discontinuous and irregular, producing jitter-like fluctuations within a narrow spatial range, often at the sub-millimeter scale. Due to the short wavelength of millimeter-wave radar (e.g., 60 GHz), such fine-scale displacements induce significant phase modulation in the radar signal, enabling the system to sensitively capture subtle jaw activity without physical contact.

\begin{figure*}[!t]
\centerline{\includegraphics[width=1\textwidth]{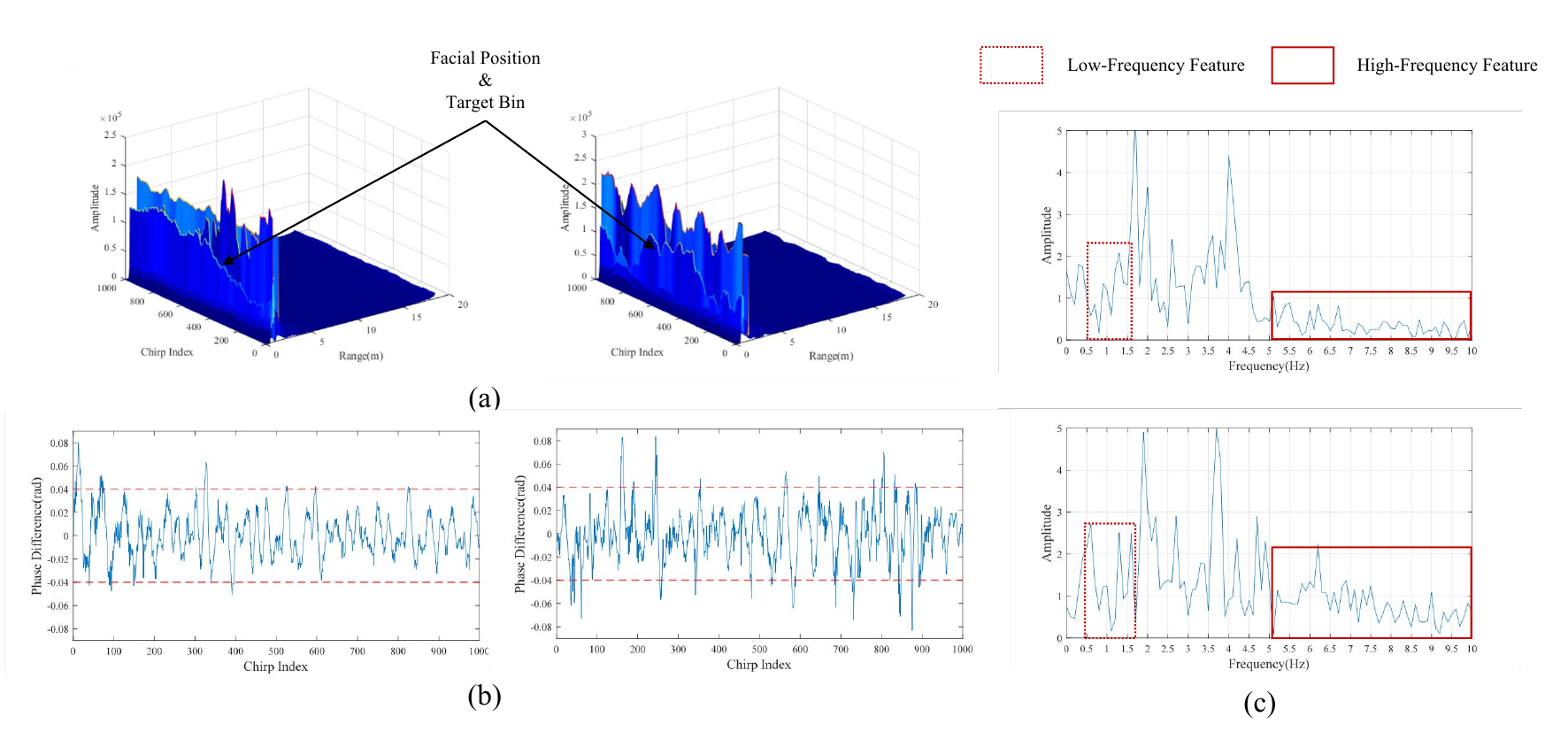}}
\caption{Comparison of The radar Signal Representation(No-grinding vs. Grinding). (a) Range-time Output obtained by 1D-FFT. (b) Time-domain phase difference on the target range bin, showing increased absolute phase difference and extrema under the influence of bruxism. (c) Enhanced energy within the 0.5–1.5 Hz band due to masseter bulging ,and increased energy within the 5–10 Hz band caused by mandibular oscillations.
.}
\label{compare}
\end{figure*}

From a signal analysis perspective, the resulting phase variations reflect complex motion dynamics in both the time and frequency domains. Temporally, the phase signal exhibits non-stationary, irregular fluctuations driven by the intermittent stick-slip behavior of the jaw. Spectrally, as shown in Fig.~\ref{compare}, these micro-movements induce broadband energy spreading, with noticeable enhancement above 5 Hz. This combination of stochastic time-domain characteristics and high-frequency spectral content forms a distinctive radar signature of bruxism, which can be leveraged for contactless recognition.

\subsection{The Effect of Masseter Bulging}
Unlike the high-frequency micro-oscillations of the mandible, the masseter muscle primarily modulates the radar echo through low-frequency surface displacement. During contraction, localized muscle thickening produces an outward shift of the skin surface, effectively varying the radar-target distance and causing phase excursions in the echo. As the muscle relaxes, this displacement reverses, yielding a quasi-periodic phase modulation corresponding to the contraction–relaxation cycle. In the time domain, such deformation appears as slow-varying baseline fluctuations superimposed on the high-frequency phase noise caused by mandibular motion. Compared with the rapid oscillations of grinding, these low-frequency variations introduce more pronounced absolute phase deviations and larger extrema within each observation window.

From a spectral viewpoint, masseter-induced deformation introduces energy concentrated in the sub-2 Hz range, typically 0.5–1.5 Hz, consistent with muscle contraction dynamics reported in prior measurements\cite{Hz}. However, this band also overlaps with involuntary body tremors and respiration-induced motion, which generate transient impulses and harmonic leakage. These interference components distort the spectral envelope, masking genuine masseter-related signatures and degrading the discriminability of frequency-domain features. Consequently, we exclude this component from spectral feature extraction and instead emphasize time-domain statistics. In particular, when masseter bulging and mandibular oscillation coexist, the resulting phase-difference sequence exhibits both broadband random fluctuations and low-frequency baseline drift. This composite modulation pattern provides a multidimensional temporal signature, capturing both high-frequency micro-vibration energy and slow deformation trends critical for reliable radar-based bruxism recognition.

While mandibular oscillations are manifested in the phase difference as an increase in variance and high-frequency energy, masseter bulging leads to a marked rise in the absolute mean and peak absolute values of the phase difference. When both effects occur simultaneously, the radar signal exhibits composite features that incorporate rapid random fluctuations along with baseline shifts. Such a composite signal pattern, characterized by both high-frequency randomness and an increase in extreme values, provides multidimensional evidence for the recognition of teeth grinding.


\begin{figure}[htbp]
\centerline{\includegraphics[width=1\linewidth]{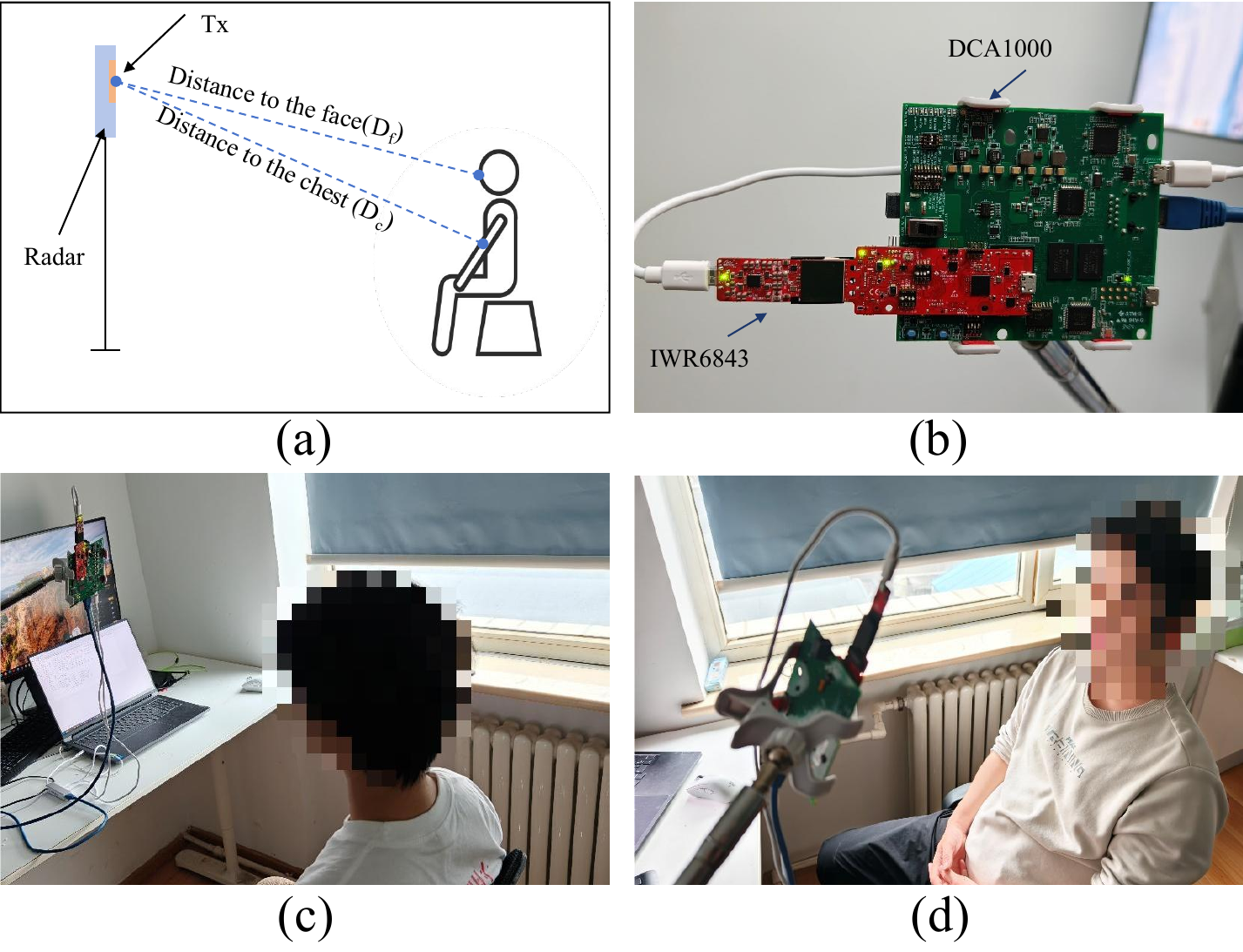}}
\caption{Experimental setup. (a) Schematic Diagram of the experimental setup. (b) Experimental equipment. (c)  Data collection.}
\label{setup}
\end{figure}
\section{Methodology}

As illustrated in Fig.~\ref{setup}, our system employs a frequency-modulated continuous-wave (FMCW) millimeter-wave radar system based on the IWR6843 development board from Texas Instruments (TI). The radar operates within the 60–64 GHz band, offering a maximum frequency modulation bandwidth of 4 GHz and a field of view of ±60° in both azimuth and elevation. The hardware configuration includes three transmit (Tx) and four receive (Rx) antennas, forming a multiple-input multiple-output (MIMO) array. However, in this system, we chose one channel of the MIMO system for the following signal processing. Intermediate frequency (IF) data from the radar chip are captured by TI’s DCA1000 high-speed data acquisition card for offline processing.

During data collection, the radar module is positioned slightly above the subject’s head and oriented perpendicularly toward the face. This configuration spatially separates facial micro-motions related to bruxism from thoracic displacements along the range dimension. The radar-to-face distance is fixed at 55 cm, which is chosen to balance the trade-off between radar sensitivity and near-field distortion. At excessively short distances, the signal will be drowned out by the DC offset caused by hardware imperfections. Meanwhile, this scenario will exacerbate near-field multipath interference and further disrupt the subjects' sleep. Conversely, at excessively long distances, the energy of the micro-motion signal will undergo attenuation, which diminishes the impact of bruxism movements on the signal and ultimately impairs recognition performance. Each recording lasts five seconds, during which participants maintain a relaxed posture and perform only teeth-grinding motions along the anterior–posterior axis. A total of 180 recordings were obtained, evenly divided between grinding and non-grinding samples.
\begin{figure}[htbp]
\centerline{\includegraphics[width=1\linewidth]{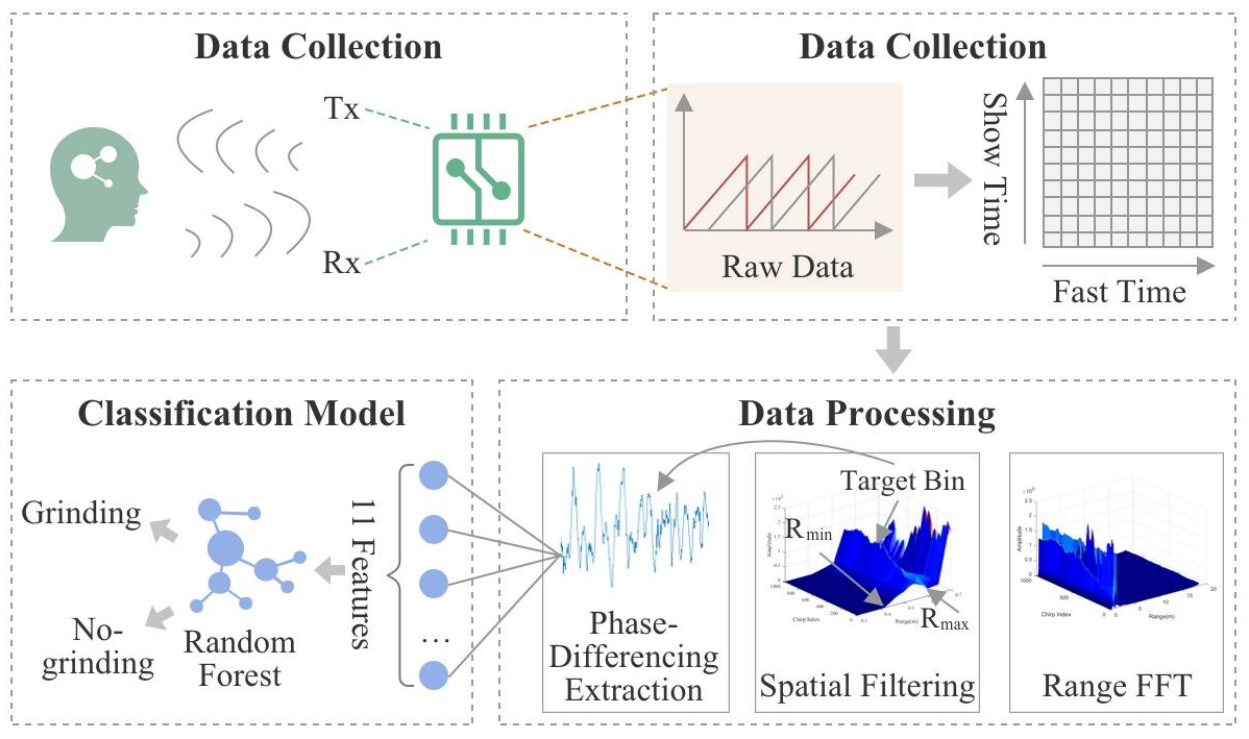}}
\caption{Workflow of contactless bruxism recognition system based on millimeter-wave radar.}
\label{frame}
\end{figure}
\subsection{Signal Preprocessing}
To extract meaningful features from raw radar echoes, acquired signals undergo a structured pre-processing workflow. The pipeline consists of three primary stages: data reorganization, spatial filtering, and phase-difference extraction.

\subsubsection{Range-Domain Reorganization}
The radar output consists of 16-bit signed integer samples representing the real and imaginary components of the intermediate frequency (IF) signal. After channel de-multiplexing, these samples are reorganized into a complex-valued IQ matrix with dimensions $[N_c, N_s]$, where $N_c$ denotes the number of chirps (slow-time dimension) and $N_s$ represents the number of samples per chirp (fast-time dimension). The fast-time dimension corresponds to the range profile within each chirp, while the slow-time dimension captures temporal variations across chirps, enabling subsequent phase-based analyses in the time domain.

\subsubsection{Spatial Filtering}
The spatial filtering stage aims to localize the radar reflections corresponding to the facial region while suppressing interference from other body parts such as the chest or shoulders. In the context of FMCW radar, this operation is implemented in the range domain through a one-dimensional Fourier transform and energy-based localization.

First, a 1D FFT is applied along the fast-time samples of the $n$-th chirp $x_n[m]$ $(m = 0, 1, \dots, N_s - 1)$:
\begin{equation}
X_n(k) = \sum_{m=0}^{N_s-1} x_n[m] \, e^{-j 2\pi \frac{km}{N_s}}, 
\quad k = 0, 1, \dots, N_s - 1,
\end{equation}
where $X_n(k)$ denotes the spectral coefficient at the $k$-th range bin for the $n$-th chirp, corresponding to a specific radial distance from the radar.

Next, incoherent integration is performed across all chirps to estimate the energy distribution along the range axis:
\begin{equation}
P(k) = \sum_{n=1}^{N_c} \big| X_n(k) \big|^2,
\quad k = k_{\text{min}}, k_{\text{min}} + 1, \dots, k_{\text{max}},
\end{equation}
where $P(k)$ represents the accumulated power at the $k$-th range bin. The indices $k_{\text{min}}$ and $k_{\text{max}}$ define the search window centered around the estimated facial distance, determined by the minimum and maximum expected ranges $R_{\text{min}}$ and $R_{\text{max}}$. Given the radar range resolution $\Delta R = \tfrac{c}{2B}$, the corresponding bin indices are:
\begin{equation}
k_{\text{min}} = \left\lfloor \frac{R_{\text{min}}}{\Delta R} \right\rfloor,
\quad
k_{\text{max}} = \left\lfloor \frac{R_{\text{max}}}{\Delta R} \right\rfloor,
\end{equation}
where $\lfloor \cdot \rfloor$ denotes the floor function.  
The target range bin is then identified by selecting the index with the maximum energy:
\begin{equation}
k^\ast = \arg \max_k P(k).
\end{equation}
This procedure effectively acts as a spatial filter that isolates the facial reflection point and suppresses returns from other range regions, ensuring that subsequent phase analysis focuses solely on bruxism-related micro-motions.

\subsubsection{Phase-Difference Extraction}
Once the facial range bin is identified, the complex phase information is extracted to capture micro-scale motion dynamics. The phase of each chirp at the selected range bin is computed as:
\begin{equation}
\phi_n = \arctan \left( \frac{\Im \{ X_n(k^\ast) \}}{\Re \{ X_n(k^\ast) \}} \right), 
\quad n = 1, 2, \dots, N_c,
\end{equation}
where $X_n(k^\ast)$ denotes the FFT output at the selected range bin.  
To maintain phase continuity and prevent wrapping within $[-\pi, \pi]$, phase unwrapping is applied to yield a temporally continuous displacement-sensitive signal.

Subsequently, the phase-difference sequence is computed:
\begin{equation}
\Delta \phi_n = \phi_n - \phi_{n-1},
\end{equation}
which functions as a temporal high-pass filter, enhancing fine-grained micro-motions associated with bruxism while attenuating low-frequency components caused by respiration or slow head movements. The resulting $\Delta \phi_n$ sequence thus serves as a motion-enhanced representation that preserves critical temporal signatures of teeth grinding behavior.

\subsection{Classification Model and Feature Selection}

Following the preprocessing and feature extraction described above, the processed radar signals are transformed into a feature vector representation suitable for supervised learning. To perform automatic bruxism recognition, a Random Forest (RF) classifier is adopted due to its strong generalization ability, resistance to overfitting, and suitability for small-sample, high-dimensional data\cite{rf}.

Given a labeled training dataset 
$\mathcal{D} = \{(\mathbf{x}_i, y_i)\}_{i=1}^N$, 
where $\mathbf{x}_i \in \mathbb{R}^d$ represents the $d$-dimensional feature vector of the $i$-th sample and $y_i$ is its corresponding label, 
the algorithm constructs $M$ decision trees $\{h_m(\mathbf{x})\}_{m=1}^M$. 
Each tree is trained on a bootstrap-resampled subset of $\mathcal{D}$, and at each node split, only a random subset of features $\mathcal{F}_m \subseteq \{1, 2, \dots, d\}$ is considered to maximize information gain:
\begin{equation}
\theta^\ast = \arg\max_{\theta \in \mathcal{F}_m} \, 
\Delta I(\theta),
\end{equation}
where $\Delta I(\theta)$ denotes the reduction in node impurity, which is typically measured by the Gini index:
\begin{equation}
I_G = 1 - \sum_{c=1}^{C} p_c^2,
\end{equation}
with $p_c$ being the proportion of class $c$ samples at a given node.
Each tree produces a class prediction $h_m(\mathbf{x}) \in \{1, 2, \dots, C\}$, and the final classification result of the ensemble is obtained via majority voting:
\begin{equation}
\hat{y} = \arg\max_{c} \sum_{m=1}^{M} 
\mathbb{I}\big(h_m(\mathbf{x}) = c\big),
\end{equation}
where $\mathbb{I}(\cdot)$ is the indicator function. This ensemble mechanism reduces variance and enhances stability compared to individual trees.

The parameters used in this study are summarized in Table~\ref{tab:rf_config}. 
The number of estimators controls the number of trees, while a 10-fold cross-validation procedure ensures generalization. The Gini impurity criterion is used for split selection, and the minimum number of samples per split is set to two.

\begin{table}[htbp]
\caption{Parameter settings for the Random Forest classifier.}
\label{tab:rf_config}
\centering
\begin{tabular}{l c}
\toprule
Parameter & Value\\
\midrule
Number of estimators & 90 \\
Cross-validation folds & 10 \\
Split criterion & Gini impurity \\
Minimum samples per split  & 2 \\
\bottomrule
\end{tabular}
\end{table}

Based on the preceding signal analysis, eleven discriminative features are extracted to describe both temporal and spectral characteristics of the radar phase-difference signals. These features include time-domain statistics (kurtosis, absolute mean, variance, and entropy), frequency-domain measures (spectral entropy, spectral variance, and energy within the 5–10~Hz band), and statistical structure descriptors (number of maxima, number of minima, number of samples greater than 0.04, and number of samples smaller than $-0.04$). 
Together, these features capture both the high-frequency stochastic behavior arising from mandibular oscillations and the low-frequency deformation effects caused by masseter bulging. 
The resulting feature space provides a compact yet comprehensive representation, enabling the Random Forest model to effectively discriminate between grinding and non-grinding patterns in the radar signal.

\begin{figure*}[!t]
\centerline{\includegraphics[width=1\textwidth]{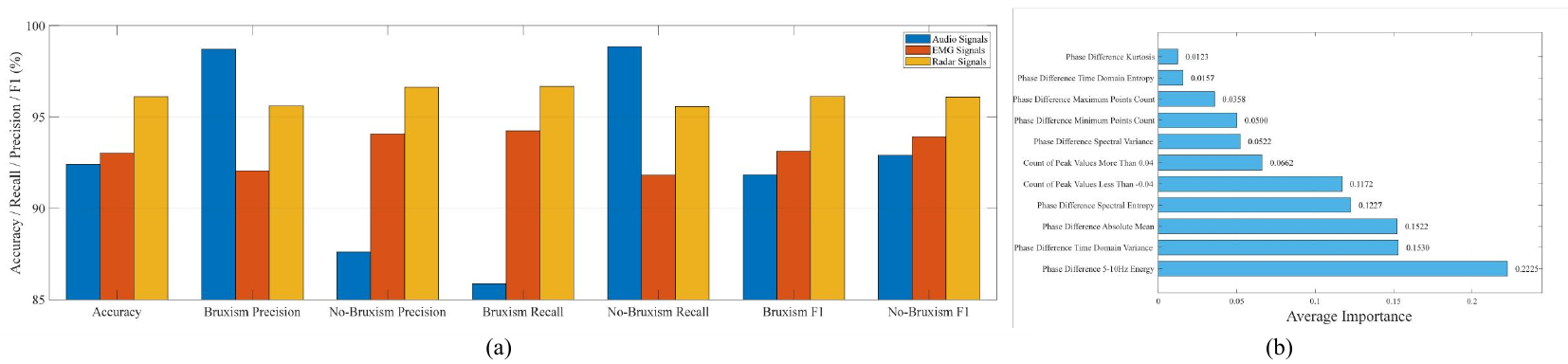}}
\caption{Comparative evaluation of different sensing modalities, and feature importance ranking. (a) Classification results of three sensing modalities in terms of accuracy, precision, recall, and F1-score. (c) Average feature importance derived from the RF classifier.}
\label{metrics}
\end{figure*}

\section{Experimental Results}

To evaluate the proposed radar-based bruxism recognition framework, experiments were conducted in a realistic indoor environment. The measurement space was an ordinary office room filled with desks, chairs, and miscellaneous objects, introducing multipath reflections and clutter typical of real-world indoor scenarios. A total of 180 five-second radar recordings were collected, including both grinding and no-grinding sequences, from three volunteers. This setup ensured diversity in individual motion patterns and verified the robustness of the system under complex indoor conditions.

Following data collection and preprocessing, the phase-difference signals were transformed into feature vectors using the methods detailed above section. These features were then used to train and validate a Random Forest classifier under a 10-fold cross-validation scheme. This evaluation protocol ensures that each sample contributes to both training and testing, thereby providing a rigorous assessment of model generalization. The classifier’s performance was quantified using four metrics: accuracy, precision, recall, and F1-score.
\begin{figure}[htbp]
\centerline{\includegraphics[width=1\linewidth]{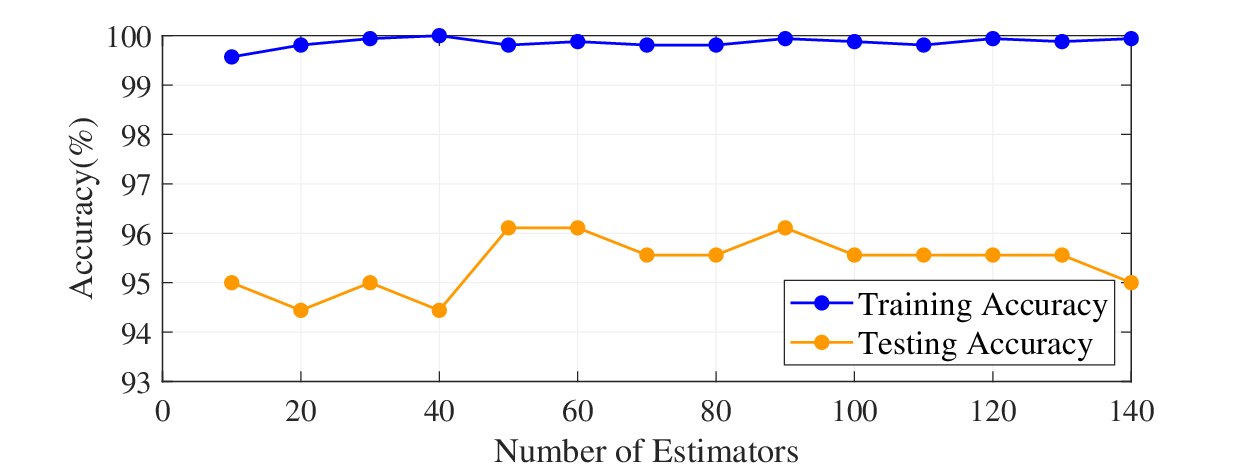}}
\caption{Accuracy performance of the RF classifier.}
\label{conf}
\end{figure}

\textbf{Accuracy:} The proportion of correctly predicted samples to the total number of samples:
\begin{equation}
Accuracy = \frac{TP + TN}{TP + TN + FP + FN},
\label{acc}
\end{equation}
where $TP$, $TN$, $FP$, and $FN$ denote true positives, true negatives, false positives, and false negatives, respectively.
The proposed method achieved 99.8\% accuracy on the training set and 96.1\% on the testing set, demonstrating strong generalization performance, as illustrated in Fig.~\ref{conf}.

\textbf{Precision:} The ratio of correctly predicted positive samples among all samples predicted as positive:
\begin{equation}
Precision = \frac{TP}{TP + FP}.
\label{pre}
\end{equation}

\textbf{Recall:} The ratio of correctly predicted positive samples among all actual positive samples:
\begin{equation}
Recall = \frac{TP}{TP + FN}.
\label{rec}
\end{equation}
Both precision and recall are crucial in bruxism recognition, as false alarms (false positives) lead to unnecessary alerts, while missed recognitions (false negatives) may overlook genuine grinding episodes.

\textbf{F1-score:} The harmonic mean of precision and recall, providing a balanced evaluation:
\begin{equation}
F1 = 2 \times \frac{Precision \times Recall}{Precision + Recall}.
\label{F1}
\end{equation}

Table~\ref{tab:radar_metrics} summarizes the quantitative results. The classifier maintains consistent performance across all metrics, confirming that the extracted features effectively capture the discriminative characteristics of bruxism-related motion.

\begin{table}[!t]
\centering
\caption{Performance Results for Radar-Based Bruxism recognition.}
\label{tab:radar_metrics}
\begin{tabular}{|l|c|c|c|c|}
\hline
\textbf{Class} & \textbf{Accuracy} & \textbf{Precision} & \textbf{Recall} & \textbf{F1-Score} \\
\hline
No-Grinding & \multirow{2}{*}{0.9611} & 0.9560 & 0.9556 & 0.9609 \\
\cline{1-1} \cline{3-5}
Grinding & & 0.9663 & 0.9667 & 0.9613 \\
\hline
\end{tabular}
\end{table}
\renewcommand{\textfloatsep}{10pt}

As shown in Fig.~\ref{metrics}(a), a comparative study was conducted using three sensing modalities—audio-based \cite{peruzzi}, EMG-based \cite{Gul}, and the proposed radar-based method. The radar approach demonstrates superior stability and accuracy across all evaluation metrics, highlighting its potential for non-contact, privacy-preserving bruxism recognition in realistic environments.
\par{}To further interpret the model, feature importance analysis was performed on the trained Random Forest classifier, as shown in Fig.~\ref{metrics}(b). The most influential features include the 5–10~Hz spectral energy, absolute mean, time-domain variance, and the number of peak values exceeding the defined thresholds. These features correspond directly to amplitude and frequency fluctuations induced by jaw muscle contractions, verifying that the proposed feature set accurately represents the physiological and motion characteristics associated with bruxism activity.

\section{Conclusion and Future Work}
This study investigated the radar-based recognition of bruxism using FMCW millimeter-wave sensing. By analyzing grinding-induced facial dynamics, specifically the combined effects of mandibular oscillations and masseter bulging, this work established the signal-level mechanisms underlying radar phase variations during teeth grinding. Based on these insights, a privacy-preserving and contactless detection framework was developed, which incorporates a spatial filtering stage, phase difference extraction, and a Random Forest classifier trained on multidimensional time-frequency characteristics. Experimental validation in a cluttered indoor environment with three volunteers and 180 grinding sessions demonstrated strong recognition ability, achieving 96.1\% testing precision, thus confirming the robustness and generalization performance of the method.

Future research will focus on extending the system to larger and more diverse populations, improving radar signal processing algorithms to enhance motion separation under complex sleeping conditions, and mitigating interference from other facial micro-movements. In addition, the integration of advanced learning architectures will be explored to further improve the robustness of classification and achieve real-time monitoring of bruxism in natural sleep environments.

\bibliographystyle{IEEEtran}  
\bibliography{reference}    

\end{document}